\documentclass[aps,prb,preprint,superscriptaddress]{revtex4}

\usepackage{graphicx}% Include figure files
\usepackage{dcolumn}% Align table columns on decimal point
\usepackage{bm}% bold math
\usepackage{longtable}

\begin{document}

%\preprint{}

\title{Magnetic, thermal and transport properties of Cd doped CeIn$_3$}


\author{N. Berry}
%\thanks{}
\affiliation{Department of Physics and Astronomy, University of
California Irvine, Irvine, CA 92697-4575}

\author{E. M. Bittar}
%\thanks{}
\affiliation{Instituto de F\'{\i}sica ``Gleb Wataghin", UNICAMP, C.
P. 6165, 13083-970, Campinas, SP, Brazil} \affiliation{Department of
Physics and Astronomy, University of California Irvine, Irvine, CA
92697-4575}

\author{C. Capan}
%\thanks{}
\affiliation{Department of Physics and Astronomy, University of
California Irvine, Irvine, CA 92697-4575}

\author{P. G. Pagliuso}
%\thanks{}
\affiliation{Instituto de F\'{\i}sica ``Gleb Wataghin", UNICAMP, C.
P. 6165, 13083-970, Campinas, SP, Brazil}

\author{Z. Fisk}
%\thanks{}
\affiliation{Department of Physics and Astronomy, University of
California Irvine, Irvine, CA 92697-4575}


\date{\today}

\begin{abstract}
We have investigated the effect of Cd substitution on the archetypal
heavy fermion antiferromagnet CeIn$_3$ via magnetic susceptibility,
specific heat and resistivity measurements. The suppression of the
Neel temperature, T$_{N}$, with Cd doping is more pronounced than
with Sn. Nevertheless, a doping induced quantum critical point does
not appear to be achievable in this system. The magnetic entropy at
$T_N$ and the temperature of the maximum in resistivity are also
systematically suppressed with Cd, while the effective moment and
the Curie-Weiss temperature in the paramagnetic state are not
affected. These results suggest that Cd locally disrupts the AFM
order on its neighboring Ce moments, without affecting the valence
of Ce. Moreover, the temperature dependence of the specific heat
below $T_N$ is not consistent with 3D magnons in pure as well as in
Cd-doped CeIn$_3$, a point that has been missed in previous
investigations of CeIn$_3$ and that has bearing on the type of
quantum criticality in this system.
\end{abstract}


\maketitle

\section{Introduction}

\begin{figure}
\includegraphics{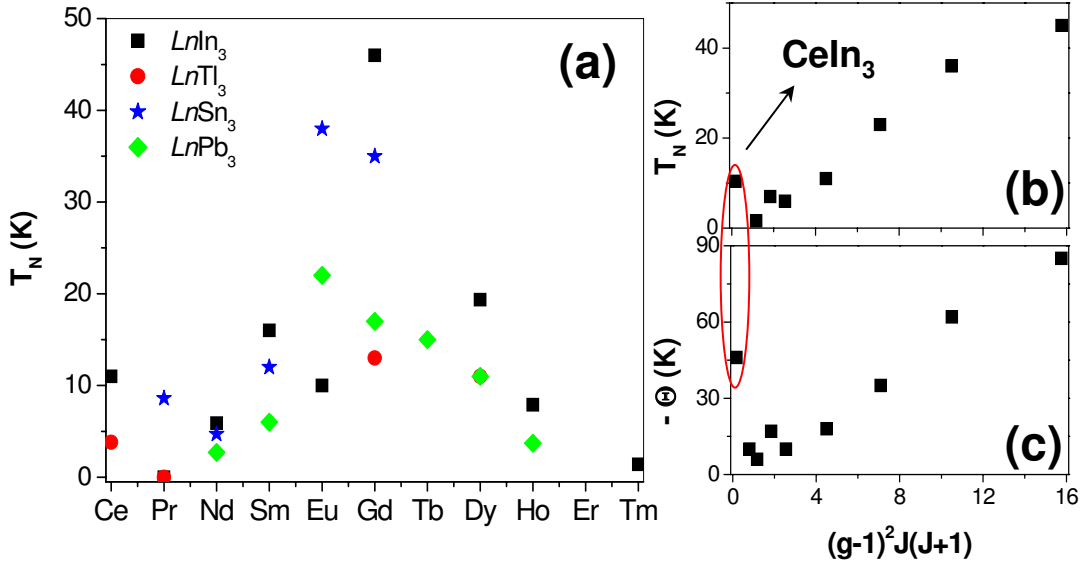}%
\caption{\label{deGennesscaling} (Color Online) (a) Neel temperature
vs rare earth in \emph{Ln}In$_3$\cite{buschow:137}, \emph{Ln}Tl$_3$,
\emph{Ln}Sn$_3$ and \emph{Ln}Pb$_3$\cite{BuschowReview}.(b),(c) Neel
temperature and Curie-Weiss temperature vs DeGennes factor
$(g^2-1)^2J(J+1)$ in \emph{Ln}In$_3$\cite{buschow:137}.}
\end{figure}

\indent CeIn$_3$ is a heavy fermion antiferromagnet (AFM) belonging
to the family of Ce-binaries that form in the cubic Cu$_3$Au
structure. Its Neel temperature $T_{N}=10.2$~K is much larger than
expected from a simple DeGennes scaling\cite{buschow:137}. Indeed
both the Neel and the Curie-Weiss temperatures for CeIn$_3$ deviate
from the DeGennes scaling, as seen in Fig.~\ref{deGennesscaling}.
The Curie-Weiss temperature in rare-earth compounds reflects the
strength of the intersite (RKKY) coupling and thus should follow the
DeGennes factor $(g^2-1)^2J(J+1)$. At low temperatures, the Kondo
screening of Ce moments is expected to reduce rather than enhance
the intersite coupling, so a larger $T_N$ is quite surprising. in
either case, this deviation might be attributed to crystal field
effects. Note that the DeGennes scaling is well-obeyed in the
related layered Ce$\emph{M}$In$_5$ and Ce$_2\emph{M}$In$_8$
($\emph{M}=$Co,Rh,Ir) compounds\cite{PhysRevB.63.054426}. CeIn$_3$
also exhibits pressure-induced superconductivity, with
$T_{c}^{max}=0.2$~K, around the critical pressure where $T_{N}$ is
suppressed to zero \cite{Mathur98, PhysRevB.65.024425}. NQR
measurements indicate homogeneous coexistence between
superconductivity and the AFM state under
pressure\cite{kawasaki:064508}. The occurrence of superconductivity
in the vicinity of the AFM quantum critical point originally lead to
the idea of magnetically mediated Cooper pairing in this and other
heavy fermion superconductors\cite{Mathur98, Monthoux07}.

\indent CeIn$_3$ is one of the most thoroughly studied systems among
heavy fermion compounds, in part due to the availability of large
single crystals of high quality. Its magnetic structure and
fluctuations are well characterized via neutron
scattering\cite{PhysRevB.22.4379, Raymond-INS}, with a commensurate
ordering wavevector of (1/2,1/2,1/2) and an ordered moment of $\sim
0.6 \mu_B$, close to the expected moment for the crystal field
ground state, $\Gamma_7$\cite{Boucherle1983409}. Evidence for Kondo
coupling has been found in resonant
photoemission\cite{PhysRevLett.46.1100, PhysRevB.56.1620}, in form
of a broad peak at the Fermi level, similar to CeSn$_3$ but with a
weaker hybridization, as well as in inelastic neutron
scattering\cite{Raymond-INS}, in form of a non-dispersive
quasi-elastic peak corresponding to a single ion Kondo scale $T_K
\sim 10$~K, of the same order as $T_{N}$. The magnetic entropy
recovered at $T_{N}$ is close to $RLn2$, meaning that the Kondo
coupling does not efficiently quench the local moments prior to
magnetic ordering. This is also supported by a moderately large
electronic specific heat coefficient in the paramagnetic state,
$\gamma \sim 180 $~mJ/K$^2$mol, corresponding to a mass enhancement
of 27 compared to the non-magnetic La-analog.

\begin{figure}
\includegraphics{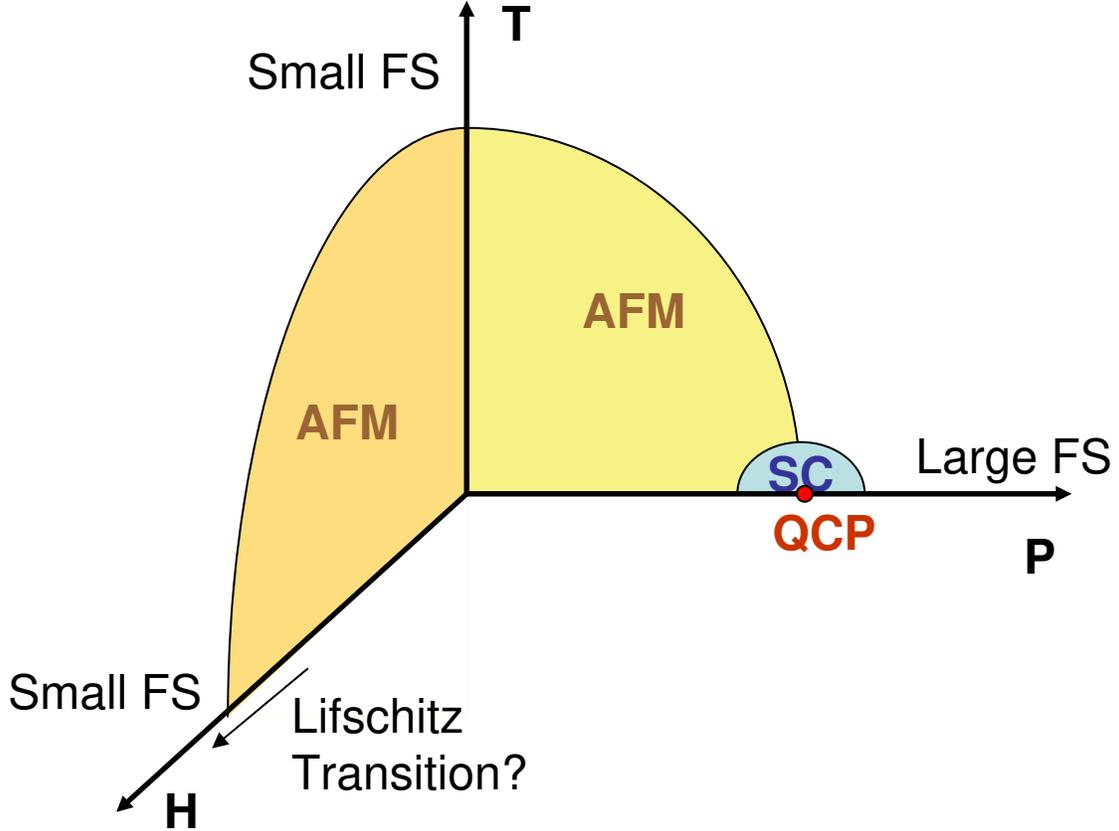}%
\caption{\label{Phasediagram} (Color Online) Schematic
temperature-pressure-magnetic field phase diagram of pure CeIn$_3$.}
\end{figure}

\indent The Fermi Surface (FS) of CeIn$_3$ continues to be a focus
of attention, in connection with the theoretical possibility of a FS
change across a zero temperature AFM instability, also called a
quantum critical point (QCP)\cite{0953-8984-13-35-202}. Hot spots
have been identified in pulsed field dHvA measurements, with a
divergent effective mass for field
$B\|[111]$\cite{PhysRevLett.93.246401}. The corresponding region of
the FS is the protruded neck in LaIn$_3$ and the mass enhancement
has been attributed to the topological change due to the AFM
Brillouin zone boundary crossing the FS\cite{gorkov:060401}. The
CeIn$_3$ Fermi surface has been mapped in the PM state via
electron-positron annihilation technique and corresponds to fully
localized $4f$-electrons\cite{PhysRevB.68.094513}. More recently it
is claimed that the divergence of the effective mass actually
happens within the Neel state, pointing to a FS topology change
generically known as a Lifshitz transition\cite{sebastian-2009}.
Moreover, the field polarized PM state (for $B \geq B_c \sim 60$~T)
has a small FS (corresponding to localized $f$-electrons) at ambient
pressure\cite{harrison:056401} but a large FS (corresponding to
itinerant $f$-electrons) at high pressure (for $p \geq
p_c$)\cite{Settai2006417, PhysRevLett.93.247003, JPSJ.74.3016}. The
phase diagram of CeIn$_3$ is schematically represented in
Fig~\ref{Phasediagram}. It is unclear how the FS continuously
evolves from small to large in the PM state with applied pressure.

\indent Evidence for a pressure induced QCP in the CeIn$_3$ phase
diagram comes from the fact that the resistivity exhibits a
temperature exponent strictly less than 2 at $p_c$, corresponding to
a breakdown of the Fermi Liquid behavior\cite{PhysRevB.65.024425}.
The FS volume increase, deduced from dHvA under
pressure\cite{Settai2006417, PhysRevLett.93.247003, JPSJ.74.3016},
across the critical pressure $p_c=2.6GPa$ where AFM is suppressed,
suggests a local QCP, where the $f$-electrons drop out of the FS
when they order magnetically\cite{0953-8984-13-35-202}. The commonly
accepted examples of a local QCP are
CeCu$_{5.9}$Au$_{0.1}$\cite{Schroder} and
YbRh$_{2}$Si$_{2}$\cite{YRS-Gegenwart}. Doping studies of CeIn$_3$,
on the other hand, are more consistent with a spin-density wave type
of QCP\cite{Hertz, Millis}, where the $f$-electrons retain their
localized character on both sides of the QCP. The AFM order in
CeIn$_{3-x}$Sn$_x$ can be suppressed to $T=0$ at $x_c\simeq0.7$,
with a logarithmically divergent electronic specific
heat\cite{Rus200562, Pedrazzini2004445}, characteristic of heavy
fermion compounds at the QCP. The divergence of the Gruneisen ratio
at this concentration has the exponent expected from a 3D-SDW
QCP\cite{kuchler:256403}. This is also supported by the fact that
there is no real breakdown of the FL behavior in the resistivity of
a $x=0.25$ sample when the Neel order is suppressed by a large
applied magnetic field\cite{silhanek:206401}. An important open
question is then: is there two distinct QCP's with different
mechanisms in this system? if so, how are the two related in a
pressure-doping phase diagram? The answer to such questions is
likely to advance significantly our
understanding of quantum criticality in heavy fermion systems.\\

\indent In this paper we report susceptibility, specific heat and
resistivity measurements in Cd doped CeIn$_3$. Cd substitution to In
is equivalent to hole doping, as opposed to electron doping with Sn,
which is known to induce a valence
transition\cite{PhysRevB.20.3770}. Cd doping in the related
Ce$\emph{M}$In$_5$ ($\emph{M}=$ Co,Rh,Ir) has lead to unexpected
results, with a few percent Cd suppressing superconductivity in
favor of the AFM state in both CeCoIn$_5$ and CeIrIn$_5$, while the
$T_N$ in CeRhIn$_5$ has a non-monotonic evolution with
doping\cite{pham:056404}. Perhaps the most noteworthy aspect is that
the effect of Cd doping can be reversed by applying
pressure\cite{pham:056404}, even though the lattice volume change
due to Cd is minute. Also, NMR measurements in Cd doped CeCoIn$_5$
suggest that Cd enhances AFM correlations locally among the
neighboring Ce ions\cite{urbano:146402}, and it remains a mystery
how such antiferromagnetic droplets can percolate at the level of a
few percent Cd introduced. These results have motivated us to
investigate the effect of Cd in CeIn$_3$ and our main findings can
be summarized as follows: Cd monotonically suppresses $T_N$, the
magnetic entropy at $T_N$, as well as the paramagnetic electronic
specific heat coefficient ($\gamma_{0}$) in CeIn$_3$, without
changing the Curie-Weiss behavior of Ce. This suggests that the
valence of Ce is not affected by Cd, in the concentration range
investigated. These results, very similar to the effect of Sn at low
doping levels, mirrors the electron-hole symmetry in the system. The
second important result is that the AFM magnon contribution to the
heat capacity is not consistent with 3D magnon spectrum, a point
that has been missed in previous reports. The possibility of 2D spin
fluctuations makes it difficult to infer the type of quantum
critical point in pure as
well as Cd doped CeIn$_3$ based on dimensional analysis.\\

\indent The paper is organized in four parts: in the first two
sections we detail the crystal growth procedure and discuss the
doping and magnetic field phase diagrams; in the following sections
we present detailed analysis of resistivity and heat capacity
measurements.\\

\section{Crystal Growth and Characterization}

\begin{table}%[H] add [H] placement to break table across pages
\caption{\label{tab:table1} Characteristic parameters of
Ce(In$_{1-x}$Cd$_x$)$_y$ single crystals: nominal ($x_{nom}$)
concentration of Cd, effective composition $x$ and $y$ as determined
from EDS, lattice constant $a$ (${\AA}$), Neel temperature $T_N(K)$,
Curie-Weiss temperature $\Theta (K)$ and effective moment $\mu_{eff}
(\mu_B)$. The concentrations noted with $\dag$ are from the two-step
growth process.}
\begin{ruledtabular}
\begin{tabular}{ccccccc}
% Lines of table here ending with \\
$x_{nom}$& $x$ & $y$ & $a$ & $T_N$ & $-\Theta$ & $\mu_{eff}$ \\
\hline
0&0&3&4.690&10.2&56.5&2.66\\
0.05&0.0198&2.63&4.690&9.5&50&2.70\\
0.1&0.0221&2.64&4.688&9.08&56.4&2.63\\
0.1$^{\dag}$&0.0174&3&--&9.8&--&--\\
0.2&0.0295&2.62&4.687&8.15&52.8&2.63\\
0.2$^{\dag}$&0.0223&3&--&9.6&--&--\\
0.3&0.0501&2.63&--&7.1&50.3&2.66\\
0.4&0.0740&2.59&4.686&6.71&49.3&2.64\\
0.4$^{\dag}$&0.0199&3&--&9.15&--&--\\
0.5&0.121&2.59&4.691&6&56.5&2.66\\
0.5$^{\dag}$&0.0199&3&--&9&--&--\\
0.6&0.0810&2.47&4.690&6.45&57.9&2.73\\
0.6$^{\dag}$&0.0303&3&--&8.8&--&--\\
\end{tabular}
\end{ruledtabular}
\end{table}

\indent Single crystals of Ce(In$_{1-x}$Cd$_{x}$)$_{y}$ were grown
out of In:Cd flux with a starting molar ratio of 1:20(1-x):20x
(Ce:In:Cd). Their characteristic parameters: composition, lattice
constant, $T_N$ and Curie-Weiss parameters are listed in
Table~\ref{tab:table1}. Energy dispersive X-ray analysis (EDS) shows
that these samples are off-stoichiometric in Indium with
$y\simeq2.6$ rather than 3. This corresponds to the surface (rather
than the bulk) composition and appears to be the result of etching
the samples in dilute HCl (in order to remove the excess In flux),
since we find the correct Ce:In ratio in the un-etched pure compound
grown on stoichiometry (1:3). From EDS, we determine the ratio of
the effective versus the nominal Cd concentration to be 1:10,
similar to the Cd doping of CeMIn$_5$ (M=Co,Rh,Ir).  We have been
unsuccessful in reaching Cd concentrations higher than effectively
$12.1\%$. Since CeCd$_3$ does not crystallize in the Cu$_3$Au
structure, excessive Cd in the flux
leads to secondary phases such as CeCd$_{11}$.\\

\indent We have also attempted to grow more stoichiometric samples
for resistivity measurements in a two-step process where we first
pre-react Ce:Cd:In in desired proportions (solid state reaction at
1180~$^{\circ}$C for 8~h in alumina crucible), followed by a second
round in the furnace with excess In flux (1:10) in the crucible
where we slowly cool (10~$^{\circ}$C/h) from 1180~$^{\circ}$C. The
resulting crystals have a ratio of Ce to (In+Cd) very close to 1:3
but are overall more dilute in Cd than the first set of crystals.
The EDS concentrations and $T_N$ (determined from $\rho$) for these
are also reported in
Table~\ref{tab:table1}.\\

\begin{figure}
\includegraphics{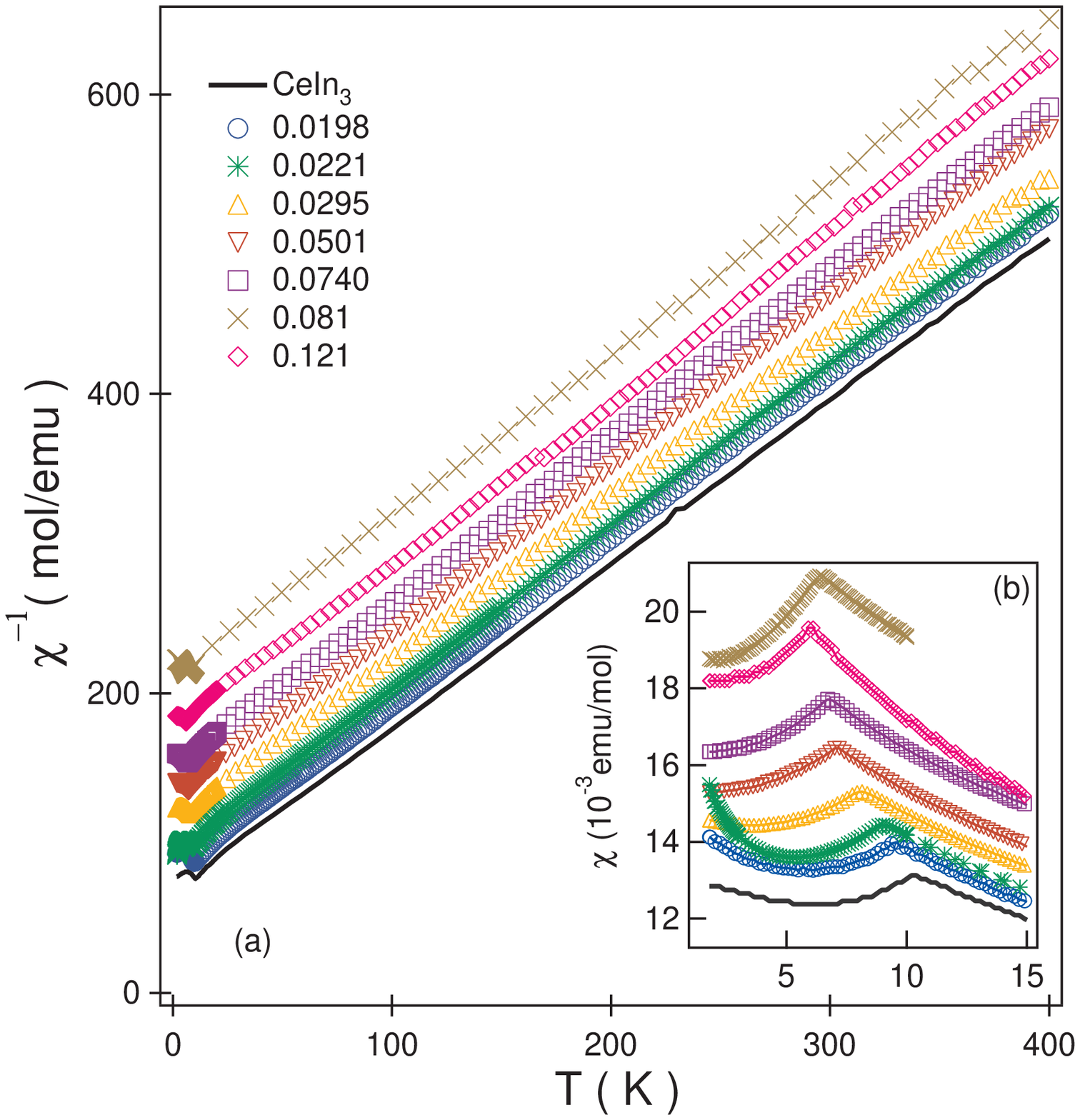}%
\caption{\label{invChi} (Color Online) (a) Inverse magnetic
susceptibility ($\chi^{-1}$) vs Temperature in
Ce(In$_{1-x}$Cd$_x$)$_{2.6}$ single crystals. The data have been
shifted vertically for clarity. The Cd concentrations as determined
from EDS are indicated. (b) Susceptibility vs Temperature for the
same samples below $T=15$~K showing a maximum at the Neel
temperature.}
\end{figure}

\indent The Curie-Weiss parameters and the Neel temperatures
reported in Table~\ref{tab:table1} are determined from the magnetic
susceptibility ($\chi$). Magnetization was measured at $H=1$~T from
$1.8$~K to $400$~K using a commercial vibrating sample SQUID
magnetometer (Quantum Design). The inverse susceptibility, shown in
Fig.~\ref{invChi}a, is linear in temperature between $100-400$~K for
all dopings. The linear fits yield a Curie-Weiss moment (slope of
$\chi^{-1}$) $\mu_{eff}$ close to the theoretical value for
Ce$^{3+}$ ($2.54\mu_B$) and an antiferromagnetic Curie-Weiss
temperature (T-axis intercept) $\Theta \simeq -50$~K, similar to the
pure compound, with no systematic variation as a function of Cd
concentration. The antiferromagnetic transition at $T_N$ is marked
by a peak in the susceptibility, as seen in Fig.~\ref{invChi}b, and
$T_N$ is monotonically suppressed with Cd doping, as listed in
Table~\ref{tab:table1} and also shown in figures~\ref{TNeel}a and \ref{TNeel}b.\\

\indent The monotonic decrease of $T_N$ with increasing Cd
concentration indicates that Cd effectively substitutes to In.
However, no systematic evolution of the lattice constant $a$ with
doping is observed (see Table~\ref{tab:table1}). The lattice
constants $a$ have been determined from Rietveld refinement of
powder X-ray diffraction spectra, using Si as a standard. One would
expect the lattice to shrink with Cd, since Cd is smaller than In. A
systematic suppression of $a$ with Cd concentration was indeed
observed in Cd doped CeCoIn$_5$\cite{booth:144519}. In addition, $a$
increases systematically with Sn in CeIn$_3$, as expected, Sn being
larger than In\cite{PhysRevB.20.3770}.\\

\indent In principle, Cd dopants could capture an electron from the
conduction band, which would force the neighboring Ce$^{3+}$ ion to
give its f electron in order to ensure electric neutrality locally,
thus becoming a non-magnetic Ce$^{4+}$ ion. Since the concentration
and effective moment of Ce$^{3+}$ ions in the PM state does not
change with Cd, this can be effectively ruled out. Thus the
Curie-Weiss analysis and the absence of change in the lattice
constant rule out any valence change of Ce induced by Cd, in the
doping range investigated. In comparison, it is known that with Sn
doping the Ce$^{3+}$ ions undergo a valence change, with CeSn$_3$
being in the (homogeneous) intermediate valence
regime\cite{PhysRevB.20.3770}. In this case, the Curie-Weiss
temperature shows a steep increase at the critical Sn concentration
where the lattice constant exhibits a kink, corresponding to the
intermediate valence regime\cite{PhysRevB.20.3770}.\\

\indent Finally, the origin of the low temperature Curie tail in the
susceptibility, observed only at low concentrations, as seen in
Fig.~\ref{invChi}b, is presently unknown and somewhat sample
dependent. A similar upturn is also present in Sn-doped
samples\cite{Pedrazzini2004445}. We have also observed such upturns
in some of the pure samples so it does not appear to be
doping induced.\\

\section{Phase diagram}

\begin{figure}
\includegraphics{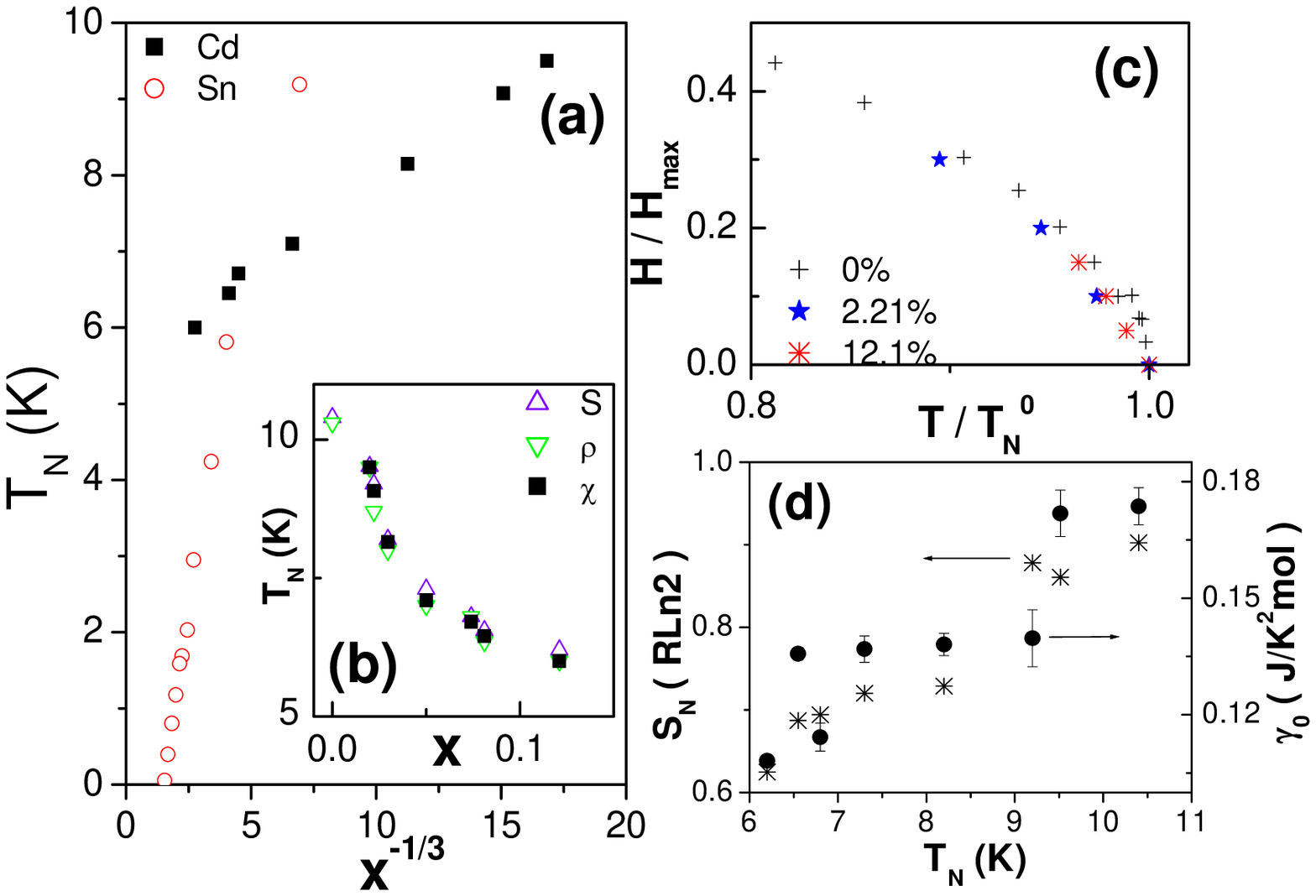}%
\caption{\label{TNeel} (Color Online) (a)  Neel Temperature ($T_N$)
vs $x^{-1/3}$ in Sn and Cd doped CeIn$_3$. The values for Sn doping
is taken from ref.\cite{Pedrazzini2004445}. $x^{-1/3}$ represents
the average distance between two dopant ions. For Cd, the effective
concentrations are used, as determined from EDS. (b) Comparison of
$T_N$, determined from different measurements, vs effective doping
$x$ in Ce(In$_{1-x}$Cd$_x$)$_{2.6}$. $T_N$ is determined from the
kink in the entropy (S), the kink in the resistivity ($\rho$) and
the maximum in the magnetic susceptibility ($\chi$). (c) Normalized
H-T phase diagram of Ce(In$_{1-x}$Cd$_x$)$_{2.6}$ for $x=0$ (+),
$2.21\%$ ($\ast$) and $12.1\%$ ($\star$). The data for pure CeIn$_3$
is from ref.\cite{PhysRevLett.93.246401}. (d) Magnetic entropy
recovered at $T_N$, $S_N$ ($\ast$) and electronic specific heat
coefficient in the PM state, $\gamma_0$ ($\bullet$) vs Neel
temperature, $T_N$ in Ce(In$_{1-x}$Cd$_x$)$_{2.6}$.}
\end{figure}

\indent The suppression of $T_N$ with Cd doping is surprising since
Cd enhances $T_N$ in the Ce$\emph{M}$In$_5$ and Ce$_2\emph{M}$In$_8$
($\emph{M}=$Co,Rh,Ir) compounds \cite{pham:056404, Adriano20093014,
adriano-2009}. This difference may be due to the fact that in the
tetragonal compounds there are two In sites and Cd preferentially
substitutes to the in-plane In\cite{booth:144519}. It is instructive
to compare the suppression of $T_N$ in Cd and Sn doped samples, as
shown in Fig.~\ref{TNeel}a. The values of $T_N$ for Sn doped samples
are from ref\cite{Pedrazzini2004445}. The values of $T_N$ in Cd
doped CeIn$_3$, determined from the peak in susceptibility, or the
kink in the entropy  and the resistivity, are in close agreement
with one and other, as shown in figure~\ref{TNeel}b. Moreover, the
consistency between the $T_N$ values for the two set of samples, In
deficient and stoichiometric, (see Table~\ref{tab:table1}) show that
the surface depletion of In does not affect the bulk properties. For
both dopants, Sn and Cd, $T_N$ follows a $x^{-1/3}$ dependence, as
shown in Fig.~\ref{TNeel}a, where $x^{-1/3}$ corresponds to the
average distance between dopants. This suggests a similar mechanisms
for the suppression of the AFM order. The weaker slope for Cd in
Fig.~\ref{TNeel}a  means a stronger suppression of $T_N$ as compared
to Sn. Nevertheless a doping induced QCP is unlikely in the case of
Cd since the x-axis intercept is negative. In the doping range
shown, neither Cd nor Sn changes the effective moment and
Curie-Weiss temperature of Ce at high temperatures, thus it is
likely that both Cd and Sn prevent their Ce neighbors from ordering
with respect to the local Weiss field below $T_N$. In other words,
the AFM order is suppressed locally around the dopants rather
than a global reduction of the RKKY coupling of Ce local moments. \\

\indent Figure~\ref{TNeel}c shows the normalized H-T phase diagram
of Cd doped CeIn$_3$, the data for the pure CeIn$_3$ is taken from
ref.\cite{PhysRevLett.93.246401}. The H-T phase diagram is
established from the heat capacity measurements at 0, 3~T, 6~T, 9~T
for $x=2.21\%$ and $12.1\%$ samples. We have verified that the
magnetic field suppression of $T_N$ is independent of x provided we
normalize $T_N$ by $T_N(H=0)$ and $H$ by $H_{max}$, where $H_{max}$
is defined as the critical field for the AFM transition at zero
temperature ($T_N(H_{max})=0$). The value of $H_{max}=60$~T is used
for pure CeIn$_3$, consistently with
Ref.\cite{PhysRevLett.93.246401}. Then $H_{max}$ is adjusted for the
Cd doped samples to give the best overlap in the normalized phase
diagram. We find $H_{max}=60$~T and $30$~T for the $x=2.21\%$
($T_N=9.075$~K) and $12.1\%$($T_N=6$~K) samples. The fact that
reducing $T_N$ by a factor of $\sim 2$ results in a suppression of
$H_{max}$ of the same rate suggests that the same effective moment
$\mu_{eff}$ is involved in the Zeeman energy ($g \mu_{eff} H_{max} =
k_B T_N$) for the $12.1\%$ Cd doped
sample as for the pure CeIn$_3$.\\

\indent Thus, the (effective) ordered moment per Ce within the AFM
state does not change with Cd, based on the H-T phase diagram. This
suggests that the CEF ground state does not change with Cd. One is
then lead to speculate that the number of Ce$^{3+}$ ions
participating in the Neel order is decreasing with increasing Cd
concentration, possibly as a result of local disruption caused by
doping, as discussed above. However, this cannot solely explain the
suppression of the AFM state since it is hard to reconcile a local
scenario with a simultaneous decrease of the magnetic entropy at
$T_N$ (see below). The investigation of the magnetic fluctuations in
the vicinity of Cd in CeIn$_3$ with NMR, of the local structure of
Ce by EXAFS as well as the investigation of possible changes in the
magnetic structure via neutron scattering
will likely shed more light on the mechanism of suppression of the AFM order with Cd.\\

\section{Heat Capacity}
\begin{figure}
\includegraphics{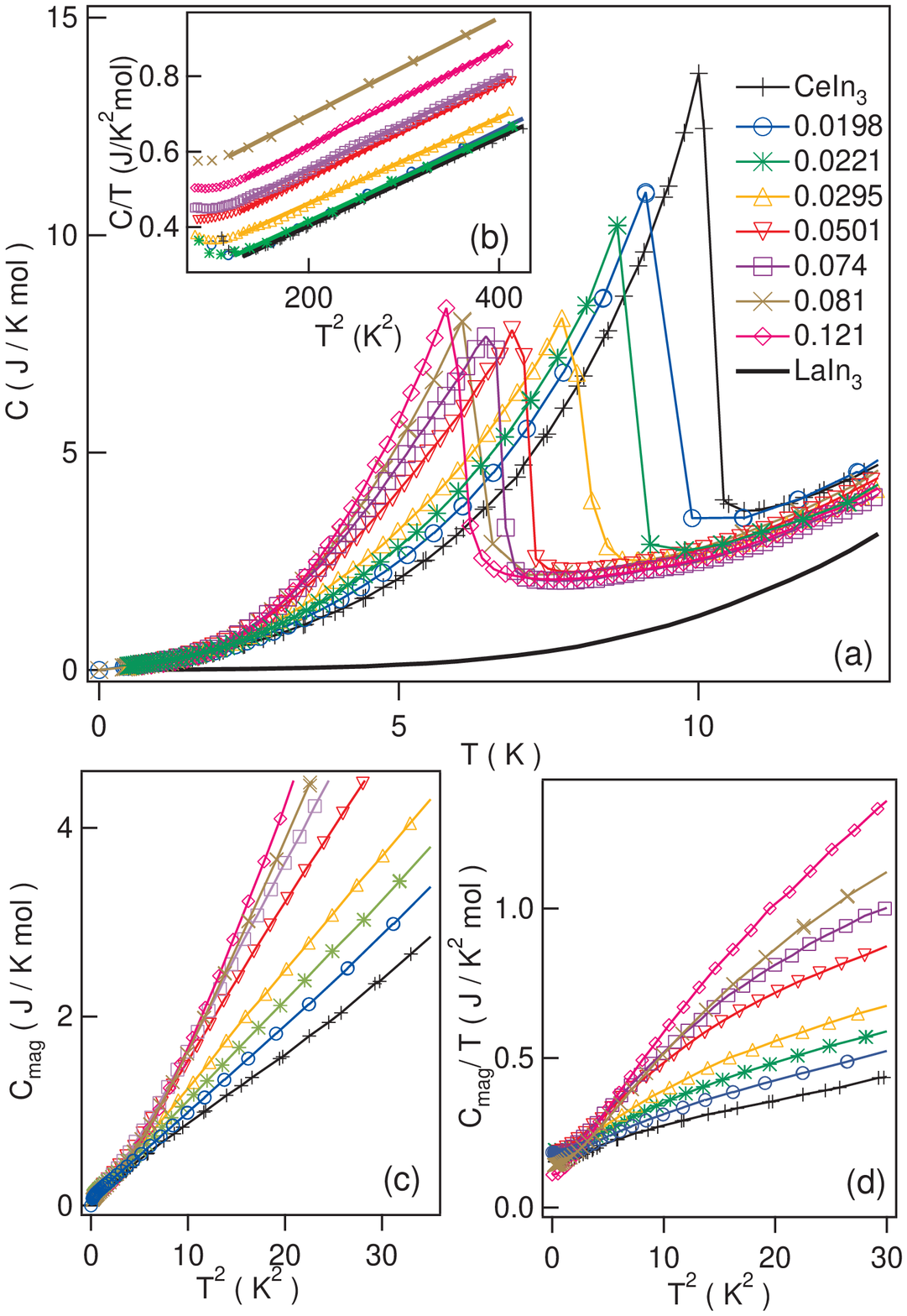}%
 \caption{\label{CvsT} (Color Online) (a) Total specific heat vs
 Temperature in single crystals of Ce(In$_{1-x}$Cd$_x$)$_{2.6}$,
 in the temperature range $0.5-15$~K, at zero magnetic field. The Cd concentrations as determined
from EDS are indicated. Also shown is the heat capacity of the
non-magnetic LaIn$_3$. (b) Specific heat divided by temperature
($\frac{C}{T}$) vs Temperature squared ($T^2$) in the paramagnetic
state for the same concentrations. Solid lines represent linear fits
to data. The data have been shifted vertically by 0.05~J/K$^2$mol
for clarity. (c) Magnetic contribution to heat capacity, $C_{mag}$,
vs Temperature squared ($T^2$) below 5~K in the same crystals. The
magnetic contribution is obtained by subtracting the lattice
contribution, as determined from LaIn$_3$. (d) $\frac{C_{mag}}{T}$
vs $T^2$ below 5~K in the same crystals.}
 \end{figure}

\subsection{The magnetic entropy at $T_N$}

\indent Figure~\ref{CvsT}a shows the temperature dependence of the
total specific heat ($C$) down to $0.5$~K in single crystals of
Ce(In$_{1-x}$Cd$_x$)$_{2.6}$ at zero magnetic field, for the
indicated nominal concentrations. The non-magnetic analog, LaIn$_3$
is also measured in order to estimate and subtract the lattice
contribution (see fig~\ref{CvsT}a). The AFM transition is marked by
a jump in $C$, characteristic of a second order transition. The
corresponding magnetic entropy $S_{mag}(T_N)=\int^{T_N}
\frac{C-C_{latt}}{T}dT$ is obtained by integrating the magnetic
contribution to specific heat up to $T_{N}$, and shown as a function
of $T_N$ in Fig.~\ref{TNeel}d. $S_{mag}$ decreases monotonically
with doping.

\indent The size of the specific heat jump is surprisingly
non-monotonic as a function of Cd concentration: it first decreases
from 0 to 2.23$\%$Cd and then increases from 2.23 to 12.1$\%$Cd.
This may correspond to the AFM transition evolving from a second
order to a weakly first order one, although the susceptibility
anomaly nor the drop in the entropy $S(T)$ at $T_N$ become
discontinuous. Further investigations of the sublattice
magnetization or the magneto-caloric effect are needed to address
this issue.

\indent The local suppression of the AFM around the Cd impurities,
as suggested above, cannot alone be responsible for the observed
entropy loss upon Cd doping. Rather, it may be due to a more
effective Kondo screening prior to ordering: if $T_N$ decreases
faster than $T_K$, the magnetic entropy at the transition will be
suppressed simply because of the quenching of Ce moments. We have
determined the single-ion Kondo temperature $T_K$ of
La$_{0.95}$Ce$_{0.05}$In$_3$ doped with 10$\%$ nominal Cd (2$\%$
effective) from specific heat (not shown) and found that the
magnetic entropy reaches $\frac{1}{2}RLn2$ at $T_K=11$~K, which
corresponds to the same single-ion Kondo temperature as in pure
CeIn$_3$\cite{Raymond-INS}. Therefore Cd has very little effect on
$T_K$ while it suppresses $T_N$ monotonically. The screening of the
Ce moments for $T_N \leq T \leq T_K$ results in a lower entropy at
$T_N$.

\subsection{The electronic specific heat coefficient}

\indent The electronic specific heat coefficient $\gamma_0$ in the
PM state is obtained from linear fits to $\frac{C}{T}$ vs $T^2$ in
the range $11-20$~K. The linearity of $\frac{C}{T}$ vs $T^2$ for all
Cd concentrations, as seen in Fig.~\ref{CvsT}b, suggests that any
additional contribution to heat capacity (from CEF excitations or
AFM fluctuations) is negligible in this range. The slight increase
in the slope of $\frac{C}{T}$ vs $T^2$ with increasing Cd
concentration results in a suppression of the T=0 intercept, which
defines $\gamma_0$. The error bars on $\gamma_0$ comes from the
choice of the temperature range used in fitting. Within the error
bars, $\gamma_0$ is suppressed with increasing Cd concentration, as
shown in Fig.~\ref{TNeel}d.

\indent The simultaneous suppression of the magnetic order and the
heavy fermion state is at odds with the Doniach phase diagram, where
a heavy paramagnetic state emerges beyond the critical point where
the AFM state is suppressed. Since the $\gamma_0$ is determined at
temperatures nominally higher than $T_K$, it may not properly
reflect the mass renormalization occurring at the lowest
temperatures (within the AFM state). Nevertheless the $\gamma_0$ in
the AFM state also appears to decrease with increased doping, as
seen in Fig.~\ref{CvsT}d, so that our data effectively rules out any
mass enhancement concomitant with the suppression of $T_N$.

\subsection{AFM magnon contribution}

\indent Figure~\ref{CvsT}c and \ref{CvsT}d show the magnetic part of
the specific heat, in the AFM state ($T < T_N$), obtained by
subtracting the lattice contribution. At low temperatures,
antiferromagnetic magnon contribution is expected to follow a $T^3$
law in 3D\cite{Kittel}. This is because the AFM magnon dispersion is
to a good approximation linear in an intermediate temperature range
and the calculation of heat capacity follows by analogy with the
Debye model. The comparison of $C_{mag}$ vs $T^2$ (Fig.~\ref{CvsT}c)
and $\frac{C_{mag}}{T}$ vs $T^2$ (Fig.~\ref{CvsT}d) reveals that the
data are more consistent with a quadratic ($T^2$) behavior rather
than cubic ($T^3$). In fact, the data collapse on a single curve at
low T for all Cd concentrations in Fig.~\ref{CvsT}c, whereas there
is no substantial T-range where $\frac{C_{mag}}{T}$  is linear in
$T^2$ in Fig.~\ref{CvsT}d.\\

\indent This unusual power law is clear evidence that the spin
fluctuations in pure as well as Cd doped CeIn$_3$ are not 3D,
contrary to what is commonly assumed for this cubic
compound\cite{kuchler:256403}. Given the ordering wavevector, one
possibility is that these are transverse spin fluctuations within
the [111] planes and the system is effectively quasi-two
dimensional. Future theoretical work, as well as a direct
investigation of magnon spectrum via inelastic neutron scattering is
strongly needed to address the origin of the $T^2$
behavior of the heat capacity in the AFM state. \\

\indent Two-dimensional fluctuations are not uncommon in the context
of local quantum criticality and they have been observed in
particular in CeCu$_{5.9}$Au$_{0.1}$ via inelastic neutron
scattering measurements\cite{Stockert}. We have checked with
previously published data\cite{Paschke} that the specific heat of
CeCu$_{5}$Au in the AFM state is also quadratic in temperature up to
0.5~K, consistently with CeIn$_3$. This gives further evidence that
in both systems the underlying physics involves 2D magnons. This
suggests that one can effectively analyze the specific heat in the
AFM state away from the QCP, to gain insight into the dimensionality
of the spin fluctuations in other quantum critical systems as well.
One immediate consequence of 2D spin fluctuations is that the sum of
the physical dimension ($d=2$) and the dynamical critical exponent
($z=2$) is exactly 4, which is the upper critical dimension in the
context of spin fluctuation theory\cite{Hertz, Millis}. In other
words, a wide fluctuation regime is possible in this system, so the
dimensional analysis\cite{0953-8984-13-35-202} alone is not
sufficient to distinguish between the SDW theory and the local QCP
scenario in this case. Then Cd doped CeIn$_3$ is a potential
candidate for FS volume change across the (pressure induced) QCP,
just as the pure compound.\\

\indent In summary, the main effect of Cd on the specific heat is:
(i) an entropy loss at the transition, (ii) a suppression of the
electronic specific heat coefficient $\gamma_0$ in the PM state.
While the former can be understood in terms of an increased Kondo
screening, the latter appears to contradict the Doniach phase
diagram. Therefore it is unlikely that Cd doping itself tunes the
system towards a QCP. Nevertheless, it is a sensible assumption
that pressure will tune Ce(In,Cd)$_3$ towards a QCP. Based on the anomalous
(quadratic) behavior of the magnon heat capacity, we speculate that the
magnetic fluctuations in this system are effectively 2D, which makes
it impractical to determine the type of QCP from dimensional analysis.\\

\section{Resistivity}

\begin{figure}
\includegraphics{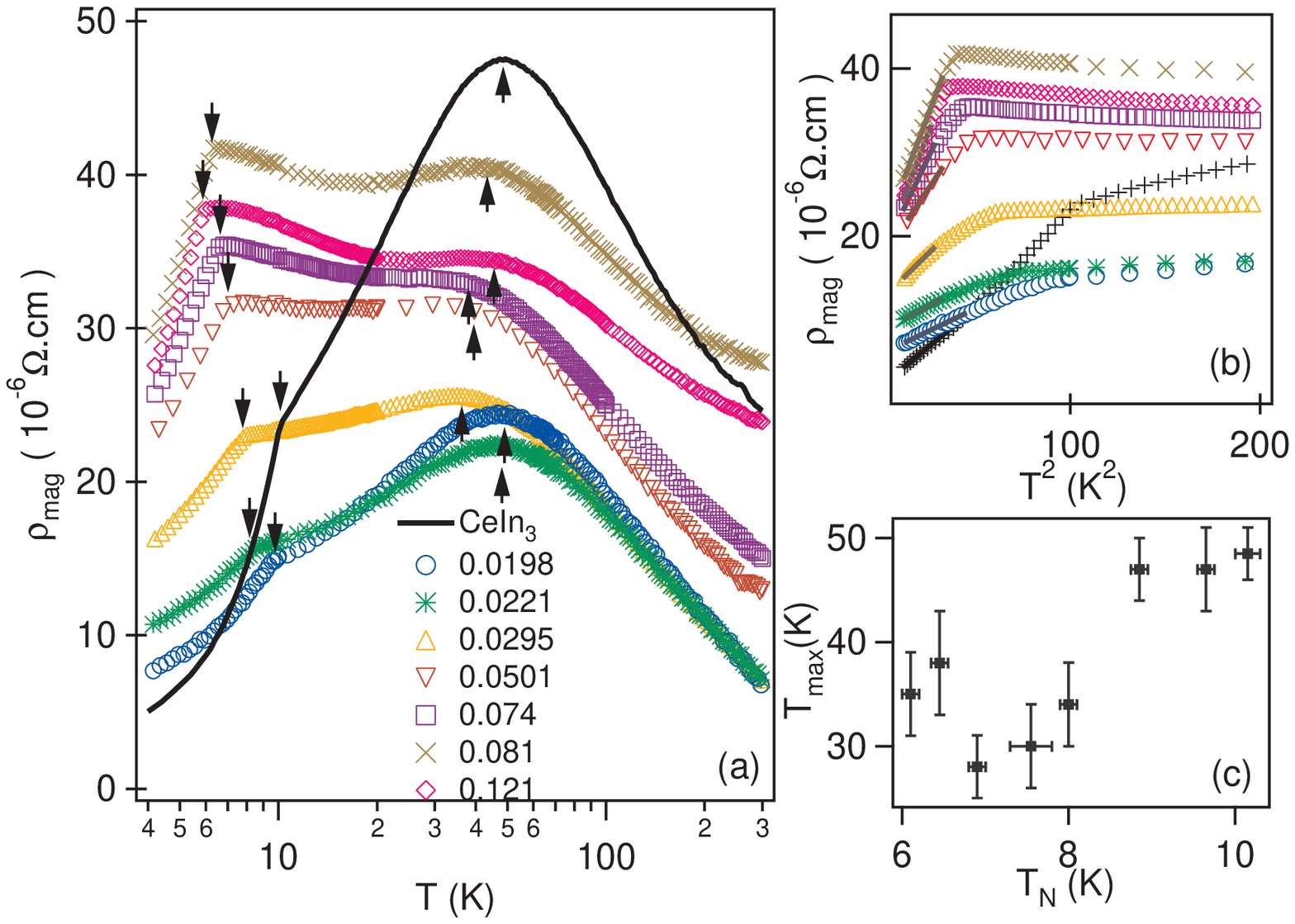}%
 \caption{\label{RhovsT1} (Color Online) (a) Magnetic contribution to the resistivity ($\rho_{mag}$) vs
Temperature in the In-deficient single crystals of
Ce(In$_{1-x}$Cd$_x$)$_{2.6}$, in the range $4-350$~K on a semi-log
scale at zero applied magnetic field. The Cd concentrations as
determined from EDS are indicated. Up  and down arrows indicate the
position of the maximum in $\rho_{mag}$ (at $T_{max}$) and of the
AFM transition (at $T_N$), respectively. The magnetic contribution
is obtained by subtracting the phonon contribution: $\rho_{mag} =
\rho - \rho_{LaIn3}$. (b) $\rho_{mag}$ vs Temperature squared
($T^2$) for the same samples. The solid lines correspond to the
Fermi Liquid fits of the form $\rho_0 + A T^2$. (c) Coherence
temperature $T_{max}$ vs Neel temperature $T_N$ with error bars,
determined graphically from $\rho_{mag}(T)$.}
\end{figure}

\begin{figure}
\includegraphics{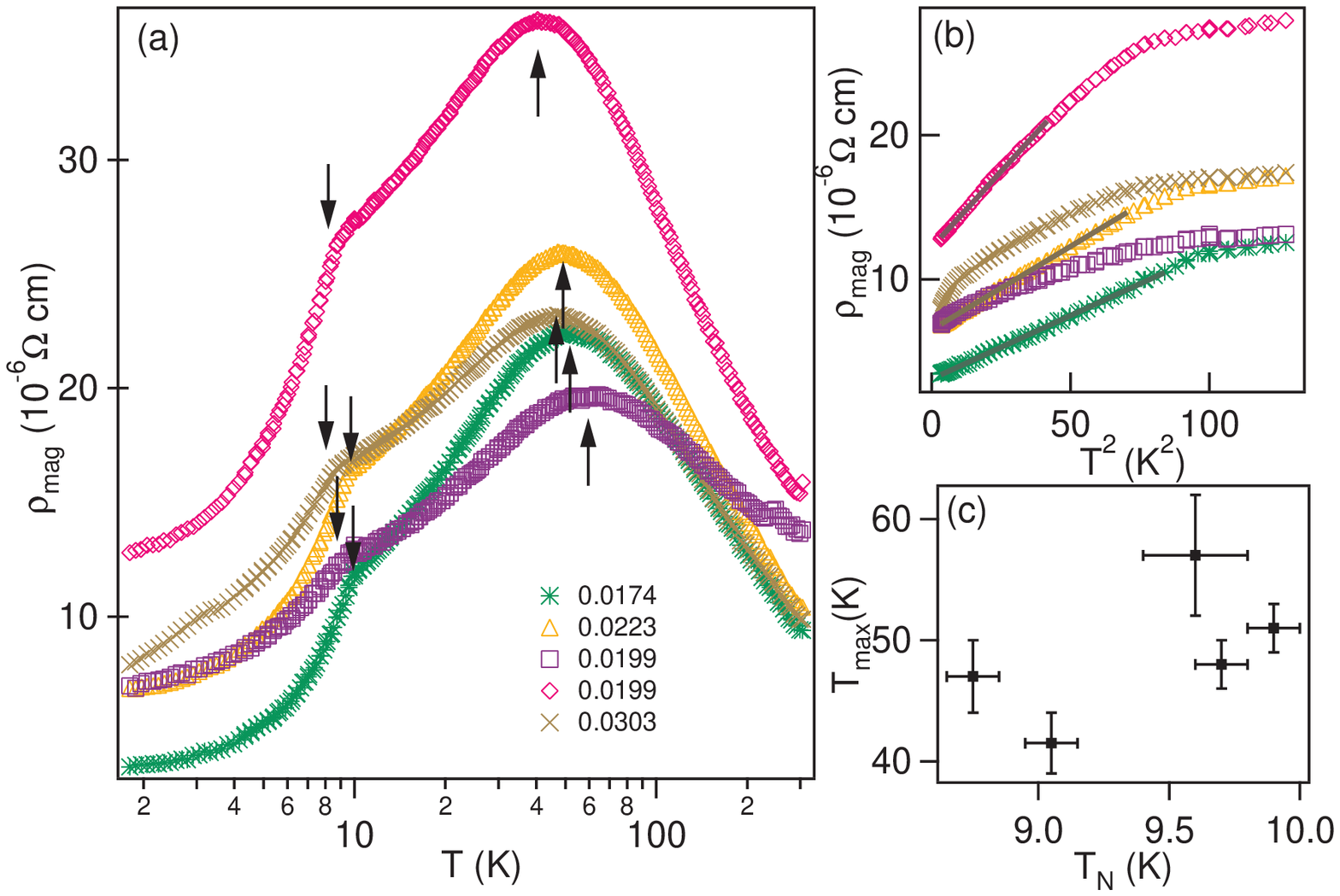}%
 \caption{\label{RhovsT2} (Color Online) (a) Magnetic contribution to the resistivity ($\rho_{mag}$) vs
Temperature in the stoichiometric single crystals of
Ce(In$_{1-x}$Cd$_x$)$_3$ between $1.8-300$~K on a semi-log scale at
$H=500$~G applied magnetic field. The Cd concentrations as
determined from EDS are indicated. Up  and down arrows indicate the
position of the maximum in $\rho_{mag}$ (at $T_{max}$) and of the
AFM transition (at $T_N$), respectively. The magnetic field ensures
that free In inclusions are in the normal state. (b) $\rho_{mag}$ vs
$T^2$ for the same samples. The solid lines correspond to the Fermi
Liquid fits of the form $\rho_0 + A T^2$. (c) Coherence temperature
$T_{max}$ vs Neel temperature $T_N$ with error bars, determined
graphically from $\rho_{mag}(T)$.}
\end{figure}

\begin{figure}
\includegraphics{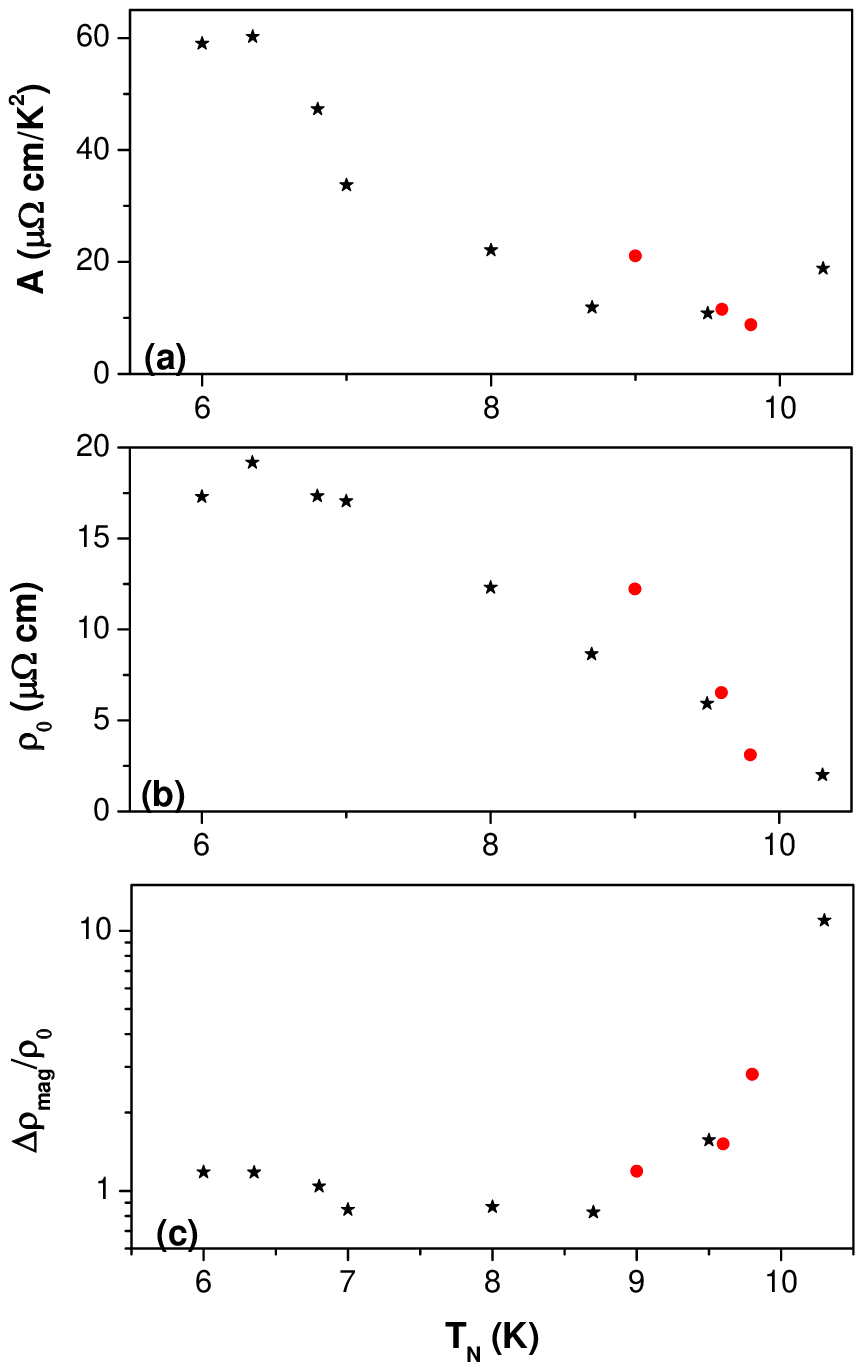}%
 \caption{\label{RhoFitParam} (Color Online)(a)Fermi Liquid $A$ coefficient
of resistivity vs Neel Temperature in the In deficient ($\star$,
$y=2.6$) and stoichiometric ($\bullet$, $y=3$) single crystals. (b)
Residual Resistivity vs Neel Temperature in the same crystals. These
parameters are obtained from fits to $\rho_{mag}$ vs $T^2$ at the
lowest temperatures, as indicated in Figures~\ref{RhovsT1} and
~\ref{RhovsT2}. (c) Normalized jump in the magnetic resistivity vs
Neel Temperature on a semi-log plot for the same samples. The jump
is defined as $\frac{\Delta \rho_{mag}}{\rho_0}=
\frac{\rho(T_N)-\rho_0}{\rho_0}$ and is due to the suppression of
spin-disorder scattering in the AFM state.}
\end{figure}

\subsection{Coherence temperature}

\indent The temperature dependence of the magnetic part of the
resistivity is shown on a semi-log plot for the two sets of Cd doped
crystals, the In deficient and the stoichiometric ones, in
Fig.~\ref{RhovsT1}a and Fig.~\ref{RhovsT2}a respectively. The
resistivity was measured with the standard four-wire technique in
the range $1.8-350$~K, with Pt wires attached to the samples using
silver paint. A current of $2$~mA was applied. For the
stoichiometric samples ($y=3$), a modest field of $500$~G was
applied in order to suppress the superconducting transition of free
In inclusions. The magnetic contribution $\rho_{mag}$ is obtained by
subtracting the phonon contribution from the total resistivity,
assuming it is the same as
for the LaIn$_3$ analog: $\rho_{mag}=\rho-\rho_{LaIn3}$.\\

\indent The characteristic peak in $\rho_{mag}(T)$ is observed at
$T_{max}=50$~K in pure CeIn$_3$, consistent with previous
reports\cite{Elenbaas1980979, PhysRevB.65.024425}. It is also known
that the peak in $\rho$ in pure CeIn$_3$ is accompanied by a
Schottky peak in the specific heat at around the same
temperature\cite{VanDiepen19711867}. This Schottky peak is
associated with a CEF excitation of $\simeq 10meV$, also seen in
inelastic neutron scattering\cite{Boucherle1983409,
PhysRevB.22.4379}. $T_{max}$ is usually taken as a crossover (or
coherence) temperature from single ion to dense Kondo regime in
heavy fermions\cite{0305-4608-12-4-015}. It also corresponds to the
crossover from a Kondo effect involving the full degeneracy of the
$J=5/2$ multiplet of the Ce$^{3+}$ ion at high $T$ to a Kondo effect
restricted to the crystal field ground state at low
$T$\cite{PhysRevB.5.4541}. The emerging picture from these two
approaches is that the Kondo lattice coherence among Ce's is only
achieved when the $f$-electrons condense into their CEF ground
state. This is also consistent with the view that the $T_{max}$ sets
the scale of intersite coupling among Ce's, a conclusion reached in
the La dilution study of CeCoIn$_5$\cite{PhysRevLett.89.106402}.

\indent In the present case, $T_{max}$ is determined graphically for
all Cd concentrations from the broad peak observed in resistivity,
as shown by the arrows in Fig.~\ref{RhovsT1}a and ~\ref{RhovsT2}a.
When plotted against $T_N$, as done in Fig.~\ref{RhovsT1}c and
~\ref{RhovsT2}c, one can see that $T_{max}$ tends to decrease with
decreasing $T_N$, due to Cd doping. A similar suppression of
$T_{max}$ is observed in Sn-doped CeIn$_3$\cite{Pedrazzini2004445}.
Note, however, that $T_{max}$ is enhanced with Sn doping in
CeCoIn$_5$\cite{bauer:245109}, highlighting the different response
of the tetragonal and cubic systems to the same dopant. The disorder
suppression of the Kondo coherence temperature in this and other
heavy fermion systems is
currently an open issue but it likely involves crystal field effects. \\

\subsection{Resistivity upturn}

\indent The most striking change induced by Cd is a clear upturn in
$\rho_{mag}(T)$ for $T_N \leq T \leq T_{max}$, as seen in
Fig.~\ref{RhovsT1}a in In-deficient samples,
Ce(In$_{1-x}$Cd$_x$)$_{2.6}$. The upturn becomes systematically more
pronounced with increasing Cd concentrations. Moreover, the
application of $H=9$~T magnetic field does not suppress the upturn
significantly (not shown). The stoichiometric compounds
(Fig.~\ref{RhovsT2}a) do not show any upturn, but this could be
simply because they are actually more dilute in Cd than their
In-deficient counterparts. Thus, In deficiency alone does not appear
to be the spurious cause of the upturn. A similar upturn is also
reported in La-doped CeIn$_3$\cite{Elenbaas1980979} as well as in
other heavy fermion systems such as Ga doped
CeAl$_2$\cite{Takeuchi1995126}. In the latter, it has been
associated with a second Kondo scale. To the best of our knowledge,
such an upturn is not found in Sn doped CeIn$_3$, nor in any Cd
doped Ce$\emph{M}$In$_5$ ($\emph{M}=$Co,Rh,Ir). Given the observed
trends, and given the absence of the upturn in the pure compound, it
is unlikely associated with a lower Kondo scale in Cd doped
CeIn$_3$. We are thus compelled to interpret it as a disorder effect
associated with
Cd doping.\\

\subsection{Spin-disorder scattering}

\indent The onset of the antiferromagnetic transition at $T_N$ is
marked by a pronounced drop in $\rho(T)$ in all Cd doped samples
(see arrows in Fig.~\ref{RhovsT1}a and ~\ref{RhovsT2}a),
corresponding to the suppression of spin-disorder scattering. In
most rare-earth intermetallics exhibiting AFM ordering, the
spin-disorder scattering in $\rho$ , as well as the ordering
temperature $T_N$, are proportional to the so-called DeGennes factor
$(g^2-1)^2J(J+1)$, since both depend on the exchange coupling
strength\cite{jensen}. Moreover, the derivative of $\rho$,
$\frac{\partial \rho}{\partial T}$ is known to mimic the jump in the
specific heat in a magnetic transition\cite{FisherLanger}, and this
is indeed the case in CeIn$_3$\cite{PhysRevB.65.024425}.

\indent In the present compounds, we found a non-monotonic evolution
with Cd of the relative change in $\rho_{mag}$ across $T_N$, namely
the ratio $\frac{\Delta\rho_{mag}}{\rho_0}$. This ratio is defined
as $\frac{\Delta \rho_{mag}}{\rho_0}=
\frac{\rho(T_N)-\rho_0}{\rho_0}$, with $\rho(T_N)$ the value of
$\rho_{mag}$ at $T_N$, and with $\rho_0$, the residual resistivity,
obtained from the quadratic fits (see below). As shown in the
semi-log plot of Fig.~\ref{RhoFitParam}c,
$\frac{\Delta\rho_{mag}}{\rho_0}$ steeply decreases upon doping,
with decreasing $T_N$, then saturates and slightly increases at
higher Cd concentrations. The latter might be a consequence of the
upturn in $\rho_{mag}(T)$ reported above. The initial sharp drop of
$\frac{\Delta\rho_{mag}}{\rho_0}$ reflects the decreasing magnetic
entropy upon doping, as expected from $\frac{\partial \rho}{\partial
T} \propto C$. This is also consistent with the pressure
results\cite{PhysRevB.65.024425}, where
the resistivity anomaly is suppressed together with $T_N$.\\

\subsection{Fermi Liquid analysis}

\indent The temperature dependence of $\rho_{mag}$ below $T_N$ is
quadratic in the pure as well as in most of the Cd doped CeIn$_3$
samples, as shown in Fig.~\ref{RhovsT1}b and \ref{RhovsT2}b. The
notable exceptions are the nominal 40$\%$ and 60$\%$ stoichiometric
samples, which exhibit a pronounced negative curvature. The
quadratic behavior of $\rho_{mag}$ is consistent with a Fermi Liquid
regime extending up to $T \simeq \frac{T_N}{2}$, also in agreement
with previous reports\cite{PhysRevB.65.024425}.

\indent The observation of a $T^2$ resistivity on a wide T-range
within the magnetically ordered state is unusual and implies a
negligible electron-magnon scattering, as compared to the
electron-electron scattering. The quadratic fits to $\rho_{mag} =
\rho_0 + AT^2$ in both the In-deficient and stoichiometric samples,
yield a Fermi Liquid coefficient $A$ and residual resistivity
$\rho_0$, both of which are shown as a function of $T_N$ on
Fig.~\ref{RhoFitParam}a and ~\ref{RhoFitParam}b. The systematic
increase in $\rho_0$ with decreasing $T_N$ (increasing Cd doping) is
simply the expected impurity scattering from Cd. The corresponding
increase in the $A$ coefficient would in principle correspond to a
mass enhancement.

\indent At first, this is surprising since the Sommerfeld
coefficient $\gamma_0$ both in the PM and the AFM states decreases
with increasing Cd concentration. In most heavy fermions, as first
noted by Kadowaki and Woods\cite{Kadowaki1986507}, the $A$
coefficient of resistivity scales with the electronic specific heat
coefficient. Therefore the thermodynamic results cast doubts on the
validity of the interpretation of the quadratic behavior of
resistivity in the AFM state in terms of Fermi Liquid behavior and
of the evolution of the $A$ coefficient with doping in terms of mass enhancement in Ce(In,Cd)$_3$.\\

\indent In summary, the effect of Cd on the resistivity is complex.
The suppression of the coherence temperature and the increase in the
residual resistivity appear to be disorder related. The smaller
spin-disorder scattering with increasing Cd concentration can be
attributed to the loss of magnetic entropy at the transition. There
is an unusual upturn induced by Cd, whose origin remains to be
elucidated. The quadratic behavior of resistivity in the AFM state
may be analyzed in terms of Fermi liquid behavior, as was previously
done in pure CeIn$_3$, however the enhancement of the $A$
coefficient is at odds with the simultaneous suppression of
$\gamma_0$ with increasing Cd concentrations.

\section{Conclusion}

\indent In conclusion, susceptibility and  resistivity measurements
in Ce(In$_{1-x}$Cd$_x$)$_3$ consistently show that the
Cd-suppression of the AFM state is not accompanied by any change of
the Ce local moments, suggesting that Cd locally disrupts the long
range order. The Sommerfeld coefficient is systematically and
significantly suppressed in the PM state, suggesting a Cd induced
suppression of the effective mass of carriers. The simultaneous
suppression of the magnetic order and the heavy fermion state is at
odds with the Doniach phase diagram and suggests that Cd tunes the
system away from the QCP. The most striking effect of Cd in the PM
state is the upturn seen in the resistivity, whose origin is
currently unknown. In the AFM state, the Fermi liquid coefficient,
determined from the resistivity, increases systematically with
increasing Cd, which would in principle imply a mass enhancement.
However, the concomitant increase in the residual resistivity as
well as the presence of magnon scattering makes this interpretation
dubious. Moreover, we found that the magnetic contribution to the
specific heat has a $T^2$ behavior in the AFM state, inconsistent
with 3D magnons. The reduced dimension for magnetic excitations,
which is the likely origin of this quadratic behavior, have been
missed in previous studies, and makes it difficult to assess the
type of quantum criticality
observed in this system based on dimensional analysis.\\

\begin{acknowledgments}
Z.F. acknowledges support through NSF Grant No. NSF-DMR-0503361.
E.M.B. and P.G.P. acknowledge support through CNPq (grant No.
200563/2008-4) and FAPESP (grant No. 2006/55347-1).
\end{acknowledgments}

%\bibliography{CeIn3_Bibtex}

\end{document}